\documentclass[12pt]{article}
\usepackage{tabls}
\usepackage{amsmath,amssymb,graphicx,multirow,subfigure,oldgerm,euscript,mathrsfs}
\usepackage[section]{placeins}
\begin{document}

\begin{titlepage}
 
\begin{flushright}
\begin{tabular}{l}
  PCCF-RI-03-09 \\
  ADP-03-135/T570 \\
  ${\rm ECT}^{*}$-03-30\\
\end{tabular}  
\end{flushright}
 
\null\vskip 0.5 true cm
%
%
\begin{center}
{{\Large \bf Simulation Methods of the Processes $\boldsymbol{B \to {\pi}^+ {\pi}^- V }$ 
 Including  $\boldsymbol{\rho^{0}-\omega}$ Mixing Effects}} \\
\vskip 1.0 cm
{ \underbar {Z.J. Ajaltouni$^1$},
O. Leitner$^{2}$,
P. Perret$^1$,
C. Rimbault$^1$,\\
A.W. Thomas$^3$} \\

\bigskip
{{\small \it $^1$ Laboratoire de Physique Corpusculaire de Clermont-Ferrand \\
IN2P3/CNRS Universit\'e Blaise Pascal \\
F-63177 Aubi\`ere Cedex France  \\
$^2$ ${\rm{ECT}}^{*}$, 
Strada delle Tabarelle, 286,\\
 38050 Villazzano (Trento), Italy \\
 $^3$ Department of Physics and Mathematical Physics and \\
 Special Research Centre for the Subatomic   Structure of Matter, \\ 
  University of Adelaide, 
  Adelaide 5005, Australia}}
\vskip 2 true cm
{\bf Talk presented at the  }
\vskip 0.8cm
{\Large {\bf CERN Workshop on Event Generators} }
\vskip 1.0cm
 {\large CERN, GENEVA, JULY 22-26, 2003}
\end{center}
\vspace{1.50cm}

\begin{abstract}

Simulation methods for the decays  $ B \to {\pi}^+ {\pi}^- V$,
where $V$ is a $1^{--}$ vector-meson, are presented in detail. 
Emphasis is put on the use of the helicity formalism and the use of 
effective Lagrangians. We show the importance of ${{\rho}^{0}}-{\omega}$ 
mixing in enhancing the direct $CP$ violation (DCPV) when the pion-pion 
invariant mass is near the mass of the $\omega$.
\end{abstract}

\end{titlepage}

\newpage

%
\section{Introduction}
%

In the framework of the LHCb experiment devoted to the search for $CP$ violation
and rare $B$ decays, special care is given to the $B$ decays into two 
vector mesons, $B \to  V_1 V_2, \ \ V_i  =  \ 1^{--}$. 
Physical motivations for studying these processes are numerous: \\

(i) Weak interaction governing the $B$ decays, the vector-mesons 
are {\it polarized} and their final states 
have specific angular distributions; which allows one to cross-check 
the Standard Model (SM) predictions and to 
perform tests of models {\it beyond} the SM. \\

(ii) In the special case of two neutral vector mesons with $\bar{V^0} = V^0$, 
the orbital angular momentum, 
$\ell$, the total spin $S$ and the $CP$ eigenvalues are related by 
the following relations: 
\begin{equation*}
\ell = \ S \ = \ 0,1,2 \  \  \Longrightarrow  CP \ = \ {(-1)}^{\ell}\ , 
\end{equation*}

\vskip 0.4cm

\noindent which implies a {\it mixing of different $CP$ eigenstates}, leading to 
$CP$ non-conservation process. 
According to Dunietz {\it et al}~\cite{Dunietzetal}, tests of $CP$ violation 
in a model independent way can 
be performed and severe constraints on models beyond the SM can be set.

%
\section{Helicity formalism and its applications}

Because the $B$ meson has spin $0$, the final two vector 
mesons, $V_1$ and $ V_2$, have the same helicity 
${\lambda}_1 = {\lambda}_2 = -1,0, +1,$ and their 
angular distribution is isotropic in the $B$ rest frame.
Let $H_w$ be the weak Hamiltonian which governs the $B$ decays. Any transition
amplitude between the initial and final states will have the following form:
\begin{equation}\label{eq1}
 H_{\lambda} = \langle V_1{(\lambda)} V_2{(\lambda)}|H_w|B \rangle\ ,
\end{equation} 
where the common helicity is ${\lambda} = -1,0, +1$.  
Then, each vector meson $V_i$ will decay into two 
pseudo-scalar mesons, $a_i,b_i$,
where  $a_i$ and $b_i$ can be either a pion or a kaon, and the angular
distributions of $a_i$ and $b_i$  depend on the $V_i$ polarization.

The helicity frame of a vector-meson, $V_i$, is defined in the $B$ 
rest frame  
such that the direction of the Z-axis is given by its momentum 
${\vec{p}_i}$. Schematically, the whole process  gets the form:
\begin{equation*}
 B     \longrightarrow     V_1  +   V_2  \longrightarrow   (a_1 + b_1)  +
 (a_2 + b_2)\ .
\end{equation*}      
The corresponding decay amplitude, $M_{\lambda}\bigl(B \rightarrow \sum_{i=1}^2
(a_i+b_i)\bigr)$, is factorized according to the relation,
\begin{equation}\label{eq2}
 M_{\lambda}\bigl(B \rightarrow \sum_{i=1}^2 (a_i+b_i)\bigr) = 
  H_{\lambda}(B \rightarrow V_1 +V_2) \times \prod_{i=1}^2 A_i(V_i \rightarrow 
a_i + b_i)\ ,    
\end{equation}
where the amplitudes $A_i(V_i \rightarrow a_i + b_i)$ are 
related to the decay of the resonances
$V_i$. The $A_i(V_i \rightarrow a_i + b_i)$  are  
given by the following expressions:
\begin{align}\label{eq3}
A_1(V_1 \rightarrow a_1 + b_1) & =   \sum_{m_1=-1}^1  c_1 
D^1_{\lambda,m_1}(0,\theta_1,0)\ ,  \nonumber  \\
A_2(V_2 \rightarrow a_2 + b_2) &  =   \sum_{m_2=-1}^1  c_2 
D^1_{\lambda,m_2}(\phi,\theta_2,-\phi)\ .  
\end{align} 
These equalities are an illustration of the Wigner-Eckart theorem. In
Eq.~(\ref{eq3}), the $c_1$
and $c_2$ coefficients represent, respectively,
the {\it dynamical  decay parameters} of the $V_1$
and $V_2$ resonances. 
The term $D^1_{\lambda,m_i}(\phi_i,\theta_i,-\phi_i)$ is the 
Wigner rotation matrix element for a spin-1 particle and we
let $\lambda{(a_i)}$ and $\lambda{(b_i)}$ be the  
respective helicities of the final particles $a_i$ and $b_i$
in the $V_i$ rest frame. $\theta_1$  is the polar angle of
$a_1$ in the $V_1$ helicity frame. 
The decay plane of $V_1$ is identified with  
the (X-Z) plane and consequently the azimuthal angle 
$\phi_1$ is set to $0$. Similarly,
$\theta_2$ and $\phi$ are respectively the polar and azimuthal angles of
particle $a_2$ in the $V_2$ helicity frame. Finally, the coefficients $m_i$
are defined as:
%
$ m_i=\lambda(a_i)-\lambda(b_i)$
%
Our convention for the $D^1_{\lambda,m_i}(\alpha,\beta,\gamma)$ 
matrix element 
is given in Rose's book~\cite{Rose}, namely: 
\begin{equation}\label{eq4}
D^1_{\lambda,m_i}(\alpha,\beta,\gamma) =  \exp[-i(\lambda \alpha +
m_i \gamma)] \;  d^1_{\lambda,m_i} (\beta)\ .
\end{equation}
The most general form of the decay amplitude ${\cal M}\bigl(B 
\rightarrow \sum_{i=1}^2
(a_i + b_i)\bigr)$ is a {\it linear superposition} 
of the previous amplitudes $M_{\lambda}\bigl(B \rightarrow \sum_{i=1}^2 (a_i
+b_i)\bigr)$ denoted by:
\begin{equation}\label{eq5}
 {\cal M}\bigl(B \rightarrow \sum_{i=1}^2 (a_i +b_i)\bigr)= 
  \sum_{\lambda} M_{\lambda}\bigl(B \rightarrow 
\sum_{i=1}^2 (a_i +b_i)\bigr)\ . 
\end{equation} 
The decay width, $\Gamma{(B \rightarrow V_1 V_2)}$, can be
computed by taking the square of the modulus,  
$\Bigl|{\cal M}\bigl(B \rightarrow \sum_{i=1}^2 (a_i +b_i)\bigr)\Bigr|$,
which involves the three kinematic parameters,  
$\theta_1, \theta_2$ and  $\phi$. This leads to the following
general expression: 
\begin{equation}\label{eq6}
d^3{\Gamma}(B \rightarrow V_1 V_2)  \propto  \Bigl|\sum_{\lambda}
M_{\lambda} \bigl( B \rightarrow
\sum_{i=1}^2 (a_i +b_i)\bigr)\Bigr|^2  =  \sum_{\lambda,\lambda'}
 h_{\lambda, \lambda'} F_{\lambda,\lambda'}(\theta_1) G_{\lambda,\lambda'}
 (\theta_2, \phi)\ ,
\end{equation} 
which involves {\bf three density-matrices},
$h_{\lambda, \lambda'}, F_{\lambda,
\lambda'}(\theta_1)$ and  $G_{\lambda, \lambda'}(\theta_2, \phi)$.
\vskip 0.4cm

\noindent $\star$ The factor $h_{\lambda, \lambda'}=H_{\lambda} H^{*}_{\lambda'}$ is an 
element of the density-matrix related to the $B$ decay. \\
 $\star$  $F_{\lambda, \lambda'}(\theta_1)$ represents the density-matrix of 
the decay $V_1 \rightarrow a_1 + b_1$. \\
$\star$  $G_{\lambda, \lambda'}(\theta_2, \phi)$ represents the
density-matrix of the decay $V_2 \rightarrow a_2+b_2$.
\par
 
\noindent The analytic expression in Eq.~(\ref{eq6}) exhibits 
a {\bf very general form}: 
 it depends on neither 
the specific nature of the intermediate resonances nor their 
decay modes (except for the spin of the final particles). 

The previous calculations are illustrated by the reaction $B^0 \rightarrow
K^{* 0} {\rho}^0$ where $K^{* 0} \rightarrow K^+ {\pi}^-$ and 
${\rho}^0 \rightarrow {\pi}^+ {\pi}^{-}$. In this channel, since all the final
particles have spin zero,  the coefficients  $m_1$ and $m_2$, defined
previously, are equal to zero. The three-fold differential
width has the following form:
\begin{multline}\label{eq7}
\frac{d^3\Gamma(B \rightarrow V_1 V_2)}{d(\cos\theta_1) 
d(\cos\theta_2) d\phi}  \propto 
\bigl(h_{++} + h_{--}\bigr){{\sin}^2{\theta_1}{\sin}^2{\theta_2}}/4 +
{h_{00}{\cos}^2{\theta_1}{\cos}^2{\theta_2}} \\
+ \Bigl\{\mathscr{R}\!e{(h_{+0})}{\cos{\phi}} - \mathscr{I}\!m{(h_{+0})}{\sin{\phi}} + \mathscr{R}\!e{(h_{0-})}{\cos{\phi}} -
\mathscr{I}\!m{(h_{0-})}{\sin{\phi}}\Bigr\}{{\sin{2\theta_1}}{\sin{2\theta_2}}}/4  \\
+ \Bigl\{\mathscr{R}\!e{(h_{+-})}{\cos{2\phi}} - \mathscr{I}\!m{(h_{+-})}{\sin{2\phi}}\Bigr\}
{{\sin}^2{\theta_1}{\sin}^2{\theta_2}}/2\ .
\end{multline}
%
It is worth noticing that the  expression in Eq.~(\ref{eq7}) is {\it
completely symmetric} in ${\theta}_1$ and ${\theta}_2$ and consequently, the
angular distribution of $a_1$ in the $V_1$ frame is {\it identical} to that of
$a_2$ in the $V_2$ frame. From Eq.~(\ref{eq7}) 
the normalized probability distribution functions (pdf) of ${\theta}_1$, 
${\theta}_2$ and $\phi$ can be derived and one finds:
%
\begin{align}\label{eq9}  
 f{({\cos}{\theta}_{1,2})} & =  {(3h_{00}-1)}{{\cos}^2{\theta}_{1,2}}  +  {(1-
 h_{00})}\ ,  \nonumber \\
 g{(\phi)} & =   1 + 2 \; \mathscr{R}\!e{(h_{+-})}{\cos{2\phi}}  -  2 \; 
\mathscr{I}\!m{(h_{+-})}{\sin{2\phi}}\ . 
\end{align}
%
%

%
%
\section{Final state interactions and $\boldsymbol{{{\rho}^0}-{\omega}}$ mixing}
%
 Hadrons produced from $B$ decays are scattered again by their mutual 
strong interactions, which 
 could modify completely their final wave-function. 
Computations of the branching ratios $B \to {\rm {Hadrons}}$ must take 
account of the
final state interactions (FSI) \cite{Quinn} which are generally divided 
into  two regimes:
 {\it perturbative} and {\it  non-perturbative}.
\par
An important question arises: how to deal with the FSI in a simple and 
practical way in order to perform
realistic and rigorous simulations?

\vskip 0.4cm
\noindent The method which has been followed for the simulations is largely 
developed in the Ref.~\cite{Ziadetal}  and is based on the hypothesis 
of {\bf Naive Factorization}, which can be summarized as
follows:
\begin{itemize}
 \item In the Feynman diagrams describing the $B$ decays into hadrons like 
{\it tree or penguin}
 diagrams, the soft gluons  exchanged among the quark lines 
are {\it neglected}.
 \item Using the Effective Hamiltonian approach and applying the 
  Operator product Expansion method (OPE), perturbative 
calculations are performed to the Wilson Coefficients (W.C.), $C_{i}$,
 at the Next to Leading Order (NLO) for an energy scale $\ge  {m_B}$.
 \item Non-perturbative effects representing physical processes at an energy 
$\le  {m_B}$ are introduced through different form-factors.
 \item The color number, $N_c$, is no longer fixed and equal to 3. It is modified according 
to the following relation:
\begin{equation*}
\frac{1}{(N_{c}^{eff})} = \frac{1}{3} + \xi\ \ ,
\end{equation*}
\end{itemize}

\noindent  where operator(s) ${\xi}$ describe(s) the non-factorizable effects.

\vskip 0.6cm

\noindent  Another important effect which appears in the channels 
$ B \to {\pi}^+ {\pi}^- V$ is the ${{\rho}^0}- {\omega}$
mixing which is an unavoidable quantum process. Indeed, the tree amplitude, 
$A^T$, and the penguin, 
$A^P$, are modified according to the following relations:
    
\begin{align}\label{eq11}
\langle K^{*} \pi^{-} \pi^{+}|H^{T}|B  \rangle & = 
\frac{g_{\rho}}{s_{\rho}s_{\omega}}
 \tilde{\Pi}_{\rho \omega}t_{\omega} +
\frac{g_{\rho}}{s_{\rho}}t_{\rho}\ , \nonumber \\
\langle  K^{*} \pi^{-} \pi^{+}|H^{P}|B  \rangle & = 
\frac{g_{\rho}}{s_{\rho}s_{\omega}} 
\tilde{\Pi}_{\rho \omega}p_{\omega} +\frac{g_{\rho}}{s_{\rho}}p_{\rho}\ .
\end{align}
Here $t_{V} \; (V=\rho \;{\rm  or} \; \omega) $ is the tree amplitude and
$p_{V}$ the penguin amplitude for 
producing a vector meson, $V$, $g_{\rho}$ is the 
coupling for $\rho^{0} \rightarrow \pi^{+}\pi^{-}$,
$\tilde{\Pi}_{\rho \omega}$ is the effective $\rho^{0}-\omega$ mixing
amplitude, and $s_{V}$ is the inverse 
propagator of the vector meson $V \ ,  
\  \ s_{V}=s-m_{V}^{2}+im_{V}\Gamma_{V} $ 
with $\sqrt s$ being the invariant mass of the $\pi^{+}\pi^{-}$ pair. 
\vskip 0.2cm 
\noindent The ratio ${A^P}/{A^T}$, which is a complex number, gets the final 
expression: 

\begin{equation}\label{eq10}
 re^{i \delta} e^{i \phi}= \frac{ \tilde {\Pi}_{\rho \omega}p_{\omega}+
s_{\omega}p_{\rho}}{\tilde 
{\Pi}_{\rho \omega} t_{\omega} + s_{\omega}t_{\rho}}\ , 
\end{equation}

\noindent where $\delta$ is the total strong phase arising both from the 
${{\rho}^0}- {\omega}$ resonance mixing and
the penguin diagram quark loop and $\phi$ is the weak angle 
resulting from the CKM matrix elements.

\section{Explicit calculations, simulations and main results}

Computations of the matrix elements are based on the effective Hamiltonian 
given by:
\begin{equation}\label{eq11a}
{\cal H}_{eff}=\frac {G_{F}}{\sqrt 2} \sum_{i} V_{CKM} C_{i}(\mu)O_i(\mu)\ ,
\end{equation}
where $G_{F}$ is the Fermi constant, $V_{CKM}$ is the 
CKM matrix element,
$C_{i}(\mu)$ are
the Wilson coefficients (W.C),
$O_i(\mu)$ are the operators related to the tree, penguin-QCD and 
penguin-EW diagrams and $\mu$ represents the renormalization scale, 
which is taken equal to $m_B$.
 
 \par
The W.C are calculated perturbatively at NLO by renormalization group 
techniques \cite{Burasetal}, while the non-perturbative parts,  
related to the operators $O_i$ and form factors. The latter  are explicitly calculated
 in the framework of the pioneering BSW models \cite{BSW}.
The free parameters which remain are: 
(i) the ratio ${q^2}/{{m_b}^2}$, where ${q^2}$ is the invariant mass 
squared of the gluon
appearing in the penguin diagram and 
(ii) the effective number of colors, ${N_c}^{eff}$. 

\vskip 0.4cm 
 
 Combining both the Wilson coefficients and the BSW formalism by including the 
${{\rho}^0}-{\omega}$ mixing in the 
$B$ meson rest-frame, where $P_B = \ {(m_B, {\vec 0})}$, 
the helicity amplitude is given by the final expression:
\begin{multline}\label{eq12} 
H_{\lambda}\bigl(B \rightarrow \rho^{0}(\omega) V_2 \bigr) = 
iB^{\rho}_\lambda(V_{ub}V_{us}^{*}c_{t_1}^{\rho}-
V_{tb}V_{ts}^{*}c_{p_1}^{\rho})+
iC^{\rho}_\lambda(V_{ub}V_{us}^{*}c_{t_2}^{\rho}-
V_{tb}V_{ts}^{*}c_{p_2}^{\rho}) +\\ 
\frac{\tilde{\Pi}_{\rho
  \omega}}{(s_{\rho}-m_{\omega}^{2})+im_{\omega}\Gamma_{\omega}}
\Bigl[ iB^{\omega}_\lambda(V_{ub}V_{us}^{*}c_{t_1}^{\omega}-
V_{tb}V_{ts}^{*}c_{p_1}^{\omega})+
iC^{\omega}_\lambda(V_{ub}V_{us}^{*}c_{t_2}^{\omega}-
V_{tb}V_{ts}^{*}c_{p_2}^{\omega})\Bigr]
\ ,
\end{multline}
where the terms $B_\lambda^{V_i}$ and $C_\lambda^{V_i}$ are combinations of 
different form factors. Their
explicit expressions, corresponding to the helicity values 
(${\lambda} = \  -1,0,+1$), are given in  Ref.~\cite{Ziadetal}.

{}From the above expression, we can deduce the dynamical 
density-matrix elements, $h_{{\lambda}, {\lambda}'}$, which 
are given by:
\begin{eqnarray*}
h_{{\lambda}, {\lambda}'} = H_{\lambda}\bigl(B \rightarrow \rho^{0}
(\omega) V_2 \bigr) H^{*}_{\lambda'}\bigl(B \rightarrow \rho^{0}
(\omega) V_2 \bigr)\ . 
\end{eqnarray*}
Because of the {\it hermiticity} of the DM, only six elements need to 
be calculated. 
\vskip 0.4cm

\subsection*{Main results}

 $\star \ h_{i,j}$ elements depend essentially on the {\bf masses} of the 
resonances; each resonance mass 
 being generated according to a {\it relativistic Breit-Wigner} distribution:
  

\begin{eqnarray*}
\frac{d\sigma}{dM^2}  =   {C_N}   \frac{\Gamma_R M_R}{{(M^2-{M^2_R})}^2 +
  {(\Gamma_R M_R)}^2}\ ,
\end{eqnarray*}
where
\begin{itemize}
 \item $C_N$ is a normalization constant.
 \item $M_R$ and $\Gamma_R$ are respectively the mass and the width of the 
vector meson.
\end{itemize}

\vskip 0.4cm

For computational reasons, the analytical treatment of 
${\rho}^0 - \omega$ mixing is simplified in the Monte-Carlo simulations \cite{Langacker}.
 
%

\noindent $\star$ The main conclusions are:

\vskip 0.3cm

 1) The spectrum of $h_{i,j}$  is too wide because of the resonance widths, 
especially the ${\rho}^0$
 width, ${\Gamma}_{\rho} = \ 150  \ {\rm MeV/c^2}. $
 
 \vskip 0.3cm
 
 2) The  longitudinal polarization, $h_{00} = \  {|H_0|}^2 \ ,$  is 
{\it largely dominant}.
 
\vskip 0.5cm 
 
  In the case of $B^0 \to \rho^0 (\omega) K^{* 0}$,   the mean value of 
$h_{00}$ is $\approx 87\%$     
while for $B^+ \to \rho^0 (\omega) \rho^{+}$, its  mean value is 
$\approx 90\%$ . 
These results have been confirmed recently by both BaBar and 
Belle collaborations~\cite{Babaretal}.

\vskip 0.3cm

3) The matrix element $h_{--} = {|H_{-1}|}^2$  is very tiny,  $\leq \  0.5\%$.

\vskip 0.5cm

4) The  non-diagonal matrix elements $h_{i,j}$ are mainly characterized by: 
\begin{itemize}
 \item  The {\bf smallness} of both their real and imaginary parts. 
 \item $  {\mathscr{I}\!m}/{\mathscr{R}\!e}  \ \ \approx  0.001 \to  0.1$. 
 \item In the special case of $B^+ \to \rho^0 (\omega) \rho^{+}, \  
{\rm \mathscr{I}\!m {(h_{i,j})}} \  \approx 0.0$.
\end{itemize} 

 \vskip 0.3cm

\noindent We arrive at the conclusion that there is a kind of  
{\bf universal behavior} of the Density-Matrix Elements,
 whatever the decay 
$B \to {\pi}^+ {\pi}^- V$ is 
($V = \  K^{*0}, K^{*{\pm}} \ , \ {\rho}^{\pm}$).



\vskip 0.4cm

\subsubsection* {Consequences for the angular distributions}

In the helicity frame of each vector-meson, $V_i$, the angular distributions 
given above (see Eq.~(8)) become simplified:

\vskip 0.4cm

$\bullet$ According to the analytic expression for $g{(\phi)}$ and 
because of the small value of $\langle h_{+-} \rangle$,
the azimuthal angle distribution is rather {\bf flat}.

\vskip 0.2cm

$\bullet$ From the expression of $f{({\cos}{\theta}_{1,2})} \ $ and 
because of the dominant longitudinal part $h_{00}$,
the polar angle distribution  is $ \approx  {{\cos}^2}{\theta}$.

\subsubsection*{Branching ratios and asymmetries}

$\star$ The energy $E_i$ and  the momentum $p_i$ of each vector meson vary 
significantly 
according to the generated event. So, the branching ratio of each channel 
must be computed by Monte-Carlo methods
from the fundamental relation:

\begin{eqnarray*}
 d{\Gamma}(B \rightarrow V_1 V_2) \ = \  {\frac{1}{8{\pi}^2 M}} \ 
{|{\cal M}(B \rightarrow V_1 V_2)|}^2 \ {\frac{d^3{\vec p_1}}{2E_1}} \ 
{\frac{d^3{\vec p_2}}{2E_2}}
 \ {\delta}^4{(P- p_1 - p_2)}\ ,
\end{eqnarray*}

\noindent and 

\begin{eqnarray*}
Br {(B \rightarrow f)}  =  \frac{\Gamma{(B \rightarrow f)}}{\Gamma{(B
    \rightarrow All)}}\ .
\end{eqnarray*}

\vskip 0.3cm

\noindent $\star$ For a fixed value of ${q^2}/{{m_b}^2}$, the BRs depend  
strongly on the Form Factor model. 
They could vary  up to a {\bf factor 2}.
\vskip 0.3cm
\noindent $\star$ The relative difference between two conjugate branching ratios, 
 $ Br{(B \rightarrow f)}$ and  
$ Br{({\bar B} \rightarrow {\bar f})}$, 
is almost  {\it independent} of the form-factor models. 

\vskip 0.3cm
 
\noindent $\star$ An interesting effect 
 is found in the variation of the 
{\bf differential asymmetry} with respect to 
the $\pi \pi$ invariant mass which is defined by:

 \begin{eqnarray*}          
 a_{CP}(m) =   \frac{{\Gamma}_m{(B \rightarrow f)} - {\bar \Gamma}_m{(\bar B
     \rightarrow {\bar f})}}{{\Gamma}_m{(B \rightarrow f)} +
 {\bar \Gamma}_m{(\bar B \rightarrow {\bar f})}}\ . 
\end{eqnarray*} 
   
\vskip 0.2cm          
  
\noindent $a_{CP}(m)$ is {\bf amplified} in the vicinity of the  $\omega$ resonance 
mass, a mass interval of
$ 20 \ {\rm MeV/}c^2$ around $ M_{\omega} = \ 782 {\rm MeV/}c^2$.

\vskip 0.2cm

This differential asymmetry is $\approx  15\%$ in the case of 
$ B^0 \to  K^{*0} {\rho}^0 {(\omega)} \ $
while it reaches $90\%$ in the channel $B^{\pm} \to  
{\rho}^{\pm} {\rho}^0 {(\omega)}$.   
It is {\it almost independent} of the form-factor model and the only 
explanation of this surprising effect is the {\bf mixing} 
of the two resonances
${\rho}^0  \ {\rm and}  \  {\omega}$ . 
These results have already been predicted analytically in the channel 
$B \to V P$ ($B \Rightarrow \rho^{0}(\omega) {\pi},  B \Rightarrow \rho^{0}(\omega) K$) by A.W. Thomas and his 
collaborators \cite{Thomas}.

\noindent Some  physical consequences can be deduced:
\begin{itemize}
 \item  A new method to detect and to measure the direct $CP$ Violation both in 
$B^0 \ {\rm and} \ B^{\pm}$ decays can be exploited.
 \item  According to the analytical expression:
\end{itemize}
 
\begin{eqnarray*}
a_{CP}^{dir} = {\frac {A^2 -{\bar A}^2}{A^2 +{\bar A}^2}} = 
\frac{-2 \ r \ {\sin {\delta}} \ {\sin {\phi}}}{1 + r^2
+2 \ r \ {\cos {\delta}} \ {\cos {\phi}}}\ , 
\end{eqnarray*}
an opportunity is offered  for measuring $\sin{\Phi}$ , where $\Phi$ is a 
{\it weak angle} resulting from
the  CKM matrix elements: \\

$\star \  \Phi = Arg{[V_{tb} V^{*}_{ts}/V_{ub} V^{*}_{us}]} \ =  \  
{\gamma} $  
 in the case of  $B \to {\rho}^0{(\omega)} K^*$, \\

$\star \  \Phi = Arg{[V_{tb} V^{*}_{td}/V_{ub} V^{*}_{ud}]} = \    
{\beta} + {\gamma} \ = {\Pi} - {\alpha} $  
in the case of  $B \to {\rho}^0{(\omega)} {\rho}^{\pm}$.
 \vskip 0.6cm
 
$\bullet$ Another interesting result deduced from the above calculations 
is the ratio ${A^P}/{A^T}$~\cite{Rimbault}.
It depends on the free parameter ${q^2}/{{m_b}^2}$ and almost independent (except in the vicinity of the $\omega$ resonance mass) 
of the ${\pi} {\pi}$ invariant mass:

\vskip 0.3cm

 \begin{center}
  ${{q^2}/{m_b^2}} = \ 0.3  \  \Longrightarrow  \  <r> \ =  0.31 \ 
{\pm} \ 0.03 $\ ,
 \vskip 0.4cm
  ${{q^2}/{m_b^2}} = \ 0.5  \  \Longrightarrow  \  <r> \ =  0.27 \ 
{\pm} \ 0.03 $\ ,
   
 \end{center}
 
 \vskip 0.4cm
 
\noindent  while the standard estimation of the ratio  ${A^P}/{A^T} \  {\rm is}  
\approx  \  30\%$  (Buras {\it et al}).
  
 \vskip 0.6cm
 
 \section{ Comparison with recent experimental results}

The Belle and BaBar collaborations recently published their first 
results concerning the charmless $B$ decays into vector mesons, $B \to VV$  \cite{Babaretal}.
 
\vskip 0.4cm

   {\underbar {Belle Collaboration}}


\begin{table}[htbp]
\begin{center}
\begin{tabular}{|l|c|c|}
\hline
 Channel&Br($\times  {10}^{-6}$)&$f_L = {|H_0|}^2$ \\
\hline
${\rho}^0 {\rho}^+$&${31.7} \pm {7.1}(stat)  ^{+3.8}_{-6.7} (syst)$ & $0.948 \pm 0.106 (stat) \pm 0.021  (syst)$    \\
\hline
 Our results &$11.0 \to 20.0$ &$90\%$ \\
\hline
\hline
\end{tabular}
\end{center}
\end{table}


{\underbar {Babar Collaboration}}



\begin{table}[htbp]
\begin{center}
\begin{tabular}{|l|c|c|c|}
\hline
 Channel&Br($\times  {10}^{-6}$)&$f_L = {|H_0|}^2$&$A_{CP}$\\
\hline
${\rho}^0 K^{*+}$&${10.6}^{+3.0}_{-2.6} {\pm} 2.4$&${0.96}^{+0.04}_{-0.15} {\pm}0.04$&${0.20}^{+0.32}_{-0.29}
{\pm} 0.04$\\
\hline
 Our results&$2.3 \to 5.8$&$87\%$&${-6.4\%}  \to   {-22\%}$\\
\hline
\hline
${\rho}^0 {\rho}^+$&${22.5}^{+5.7}_{-5.4} {\pm} 5.8$&${0.97}^{+0.03}_{-0.07} {\pm}0.04$&${-0.19} {\pm}0.23
{\pm} 0.03$\\
\hline
 Our results&$11.0 \to 20.0$&$90\%$&${-8.5\%}  \to   {-10\%}$\\
\hline
\hline
\end{tabular}
\end{center}
\end{table}

  
 \vskip 0.6cm
 
Because there is as yet insufficient data to allow one to bin data in the 
region of the $\omega$ resonance, one can only look at the global asymmetry 
$A_{CP}$ measured by BaBar. This is compatible with zero and 
the differential asymmetry with regard to ${\pi} {\pi}$ has not been investigated.
 We look forward with great anticipation to the time 
 when the invariant mass distribution can be investigated.

\section{Conclusion and perspectives}

Monte-Carlo methods based on the helicity formalism have been used for all the numerical simulations of
the channels $B \to {\pi}^+ {\pi}^- V$  with  $V = \ K^{*0}, K^{*{\pm}}, {\rho}^{\pm}$. Rigorous and
detailed calculations of the $B$ decay density-matrix have been carried out completely and the corresponding
code has been already implemented in the {\bf LHCb} generator code.

$\bullet$  Despite the fact that the naive factorization hypothesis is 
very useful for 
weak hadronic $B$ decays, this method is limited because it involves 
theoretical  uncertainties, some of them 
being very large.

\vskip 0.4cm

\noindent However the physical consequences of this study are very interesting~:
\begin {itemize}
 \item The {\it form factor model} plays important role, especially in the 
estimation of the different
 branching ratios which can vary by up to a factor of 2.
 \item The longitudinal polarization is largely dominant, 
whatever the form factor model.
 \item The ${{\rho}^0}- {\omega}$ mixing is the main ingredient in 
the enhancement of direct $CP$ violation.
 \item  A new way to look for $CP$ Violation is found and it can help to 
develop new methods  for measuring the  angles 
{\bf ${\gamma}  \  {\rm and} \  {\alpha} $}.
\end{itemize}


What remains to be done is to cross-check these predictions with experimental data coming from the
LHC experiments.

\subsubsection*{Acknowledgments}
One of us, Z.J.A., is very indebted to the organizers of this workshop, 
Dr Nick Brook and Dr Witek Pokorski
from LHCb collaboration, 
for the opportunity they gave him to present the recent study on 
$B \to VV$ made 
in collaboration with theoretical colleagues at Adelaide University.
\par
 All of us enjoyed  
the exciting and illuminating discussions we got during the parallel sessions
and regarding the broader aspects of $B$ physics.

\newpage

%
%
%
%
%
%
%

%
\newpage



\newpage
%

\begin{figure}
\centering\includegraphics[height=20.0cm, width = 15.0cm,clip=true]{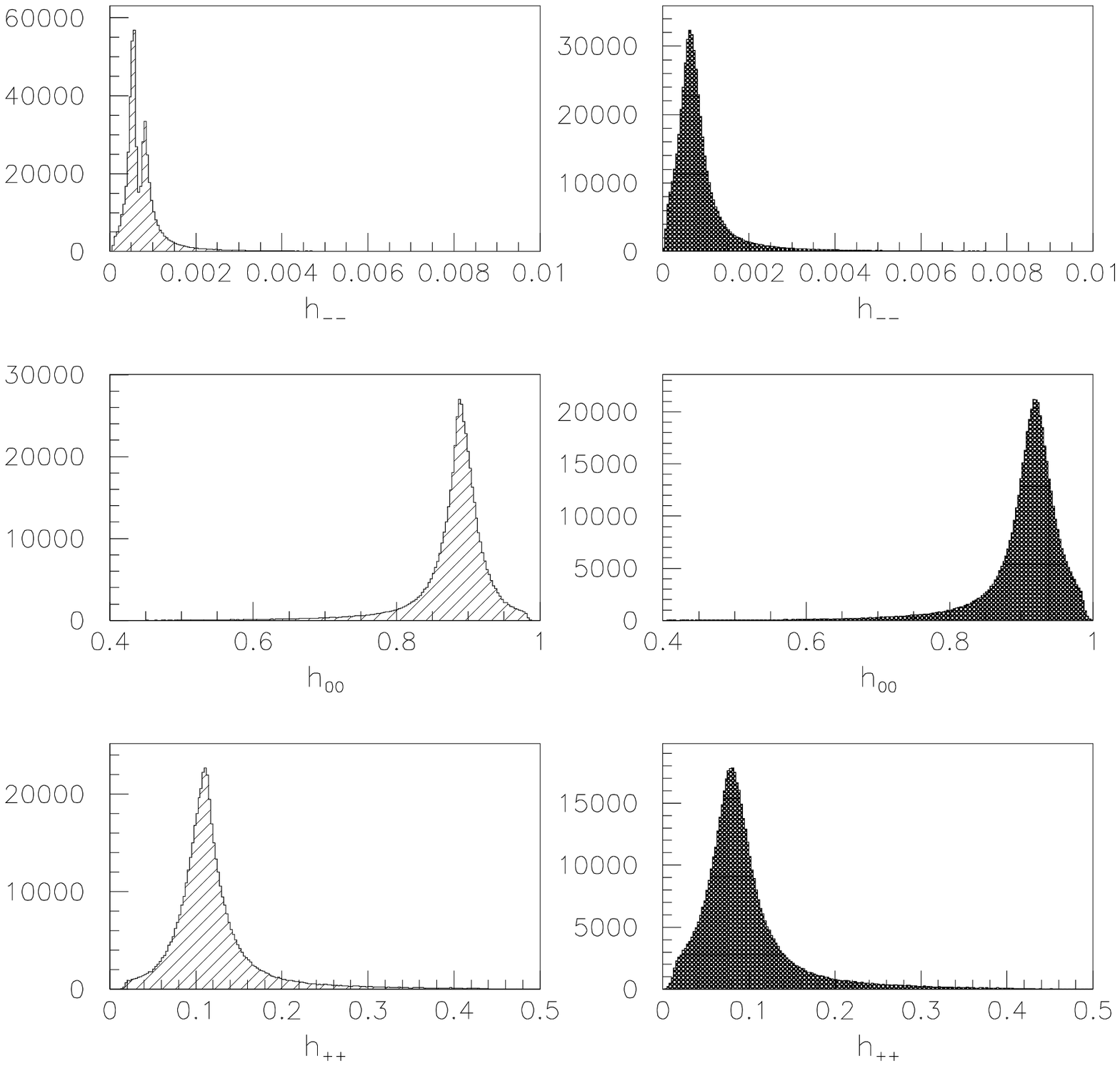}
\caption{Spectrum of $ h_{--}, h_{00}, h_{++}$. Histograms on the  left
  correspond to the channel $B^0 \rightarrow \rho^{0}(\omega) K^{* 0}$
  where the parameters used 
  are: ${q^2}/{{m^2}_b} =  0.3$, $N_c^{eff} = 2.84$, $\rho= 0.229,
  \eta=0.325$ and form factors from the GH model. 
  Histograms on the right correspond
  to the channel $B^+ \rightarrow \rho^{0}(\omega) \rho^+$ for the same
  parameters with $N_c^{eff} = 2.01$.}
\label{fig1}
\end{figure}

\begin{figure}
\centering\includegraphics[height=20.0cm, width = 15.0cm,clip=true]{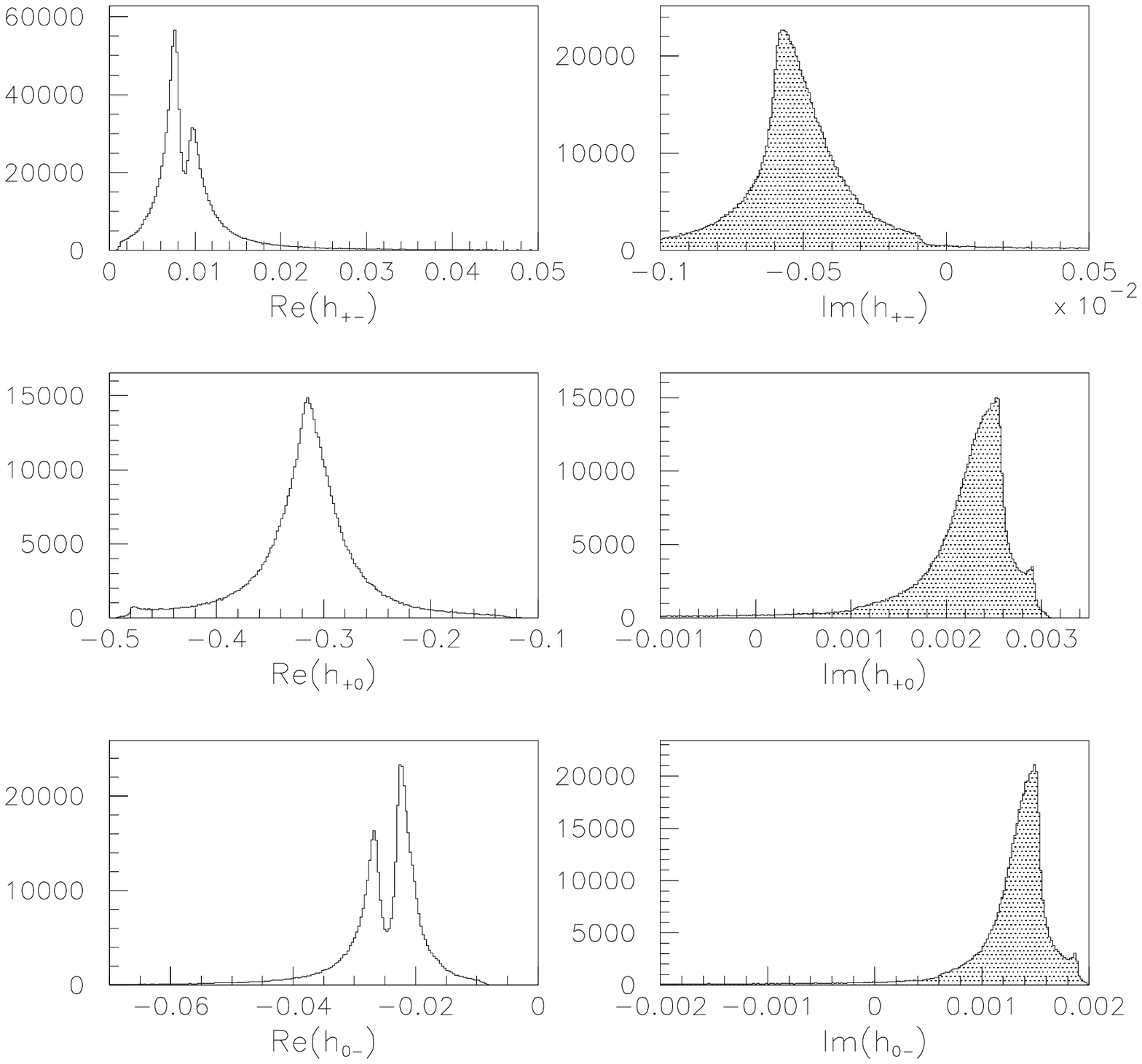}
\caption{Spectrum of $ \mathscr{R}\!e{(h_{ij})} \;  {\rm  and} \;  \mathscr{I}\!m{(h_{ij})}$
  where $i \neq j$.   Histograms 
  correspond to channel $B^0 \rightarrow \rho^{0}(\omega) K^{* 0}$ where
  the used 
  parameters are: ${q^2}/{{m^2}_b} =  0.3$, $N_c^{eff} = 2.84$, $\rho=0.229,
  \eta=0.325$ and form factors from the GH model. }
\label{fig2}
\end{figure}
\begin{figure}
\centering\includegraphics[height=20.0cm, width = 15.0cm,clip=true]{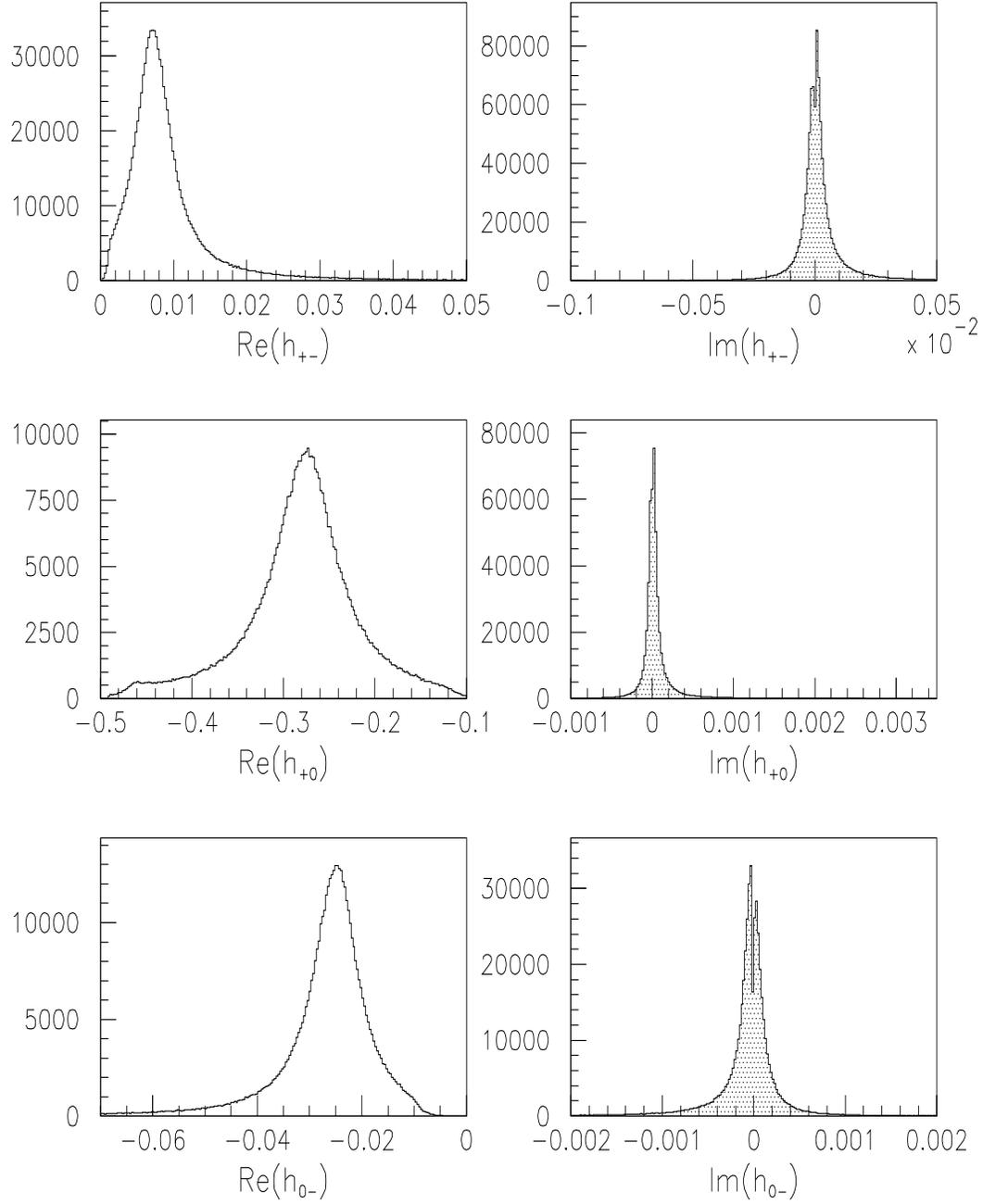}
\caption{Spectrum of $ \mathscr{R}\!e{(h_{ij})} \;  {\rm  and} \;  \mathscr{I}\!m{(h_{ij})}$
  where $i \neq j$.   Histograms 
  correspond to the channel $B^+ \rightarrow \rho^{0}(\omega) \rho^+$ where
  the used
  parameters  are: ${q^2}/{{m^2}_b} =  0.3$, $N_c^{eff} = 2.01$, $\rho=0.229,
  \eta=0.325$ and form factors from the GH model.}
\label{fig3}
\end{figure}
\begin{figure}
\centering\includegraphics[height=9.0cm, width = 8.0cm,clip=true]{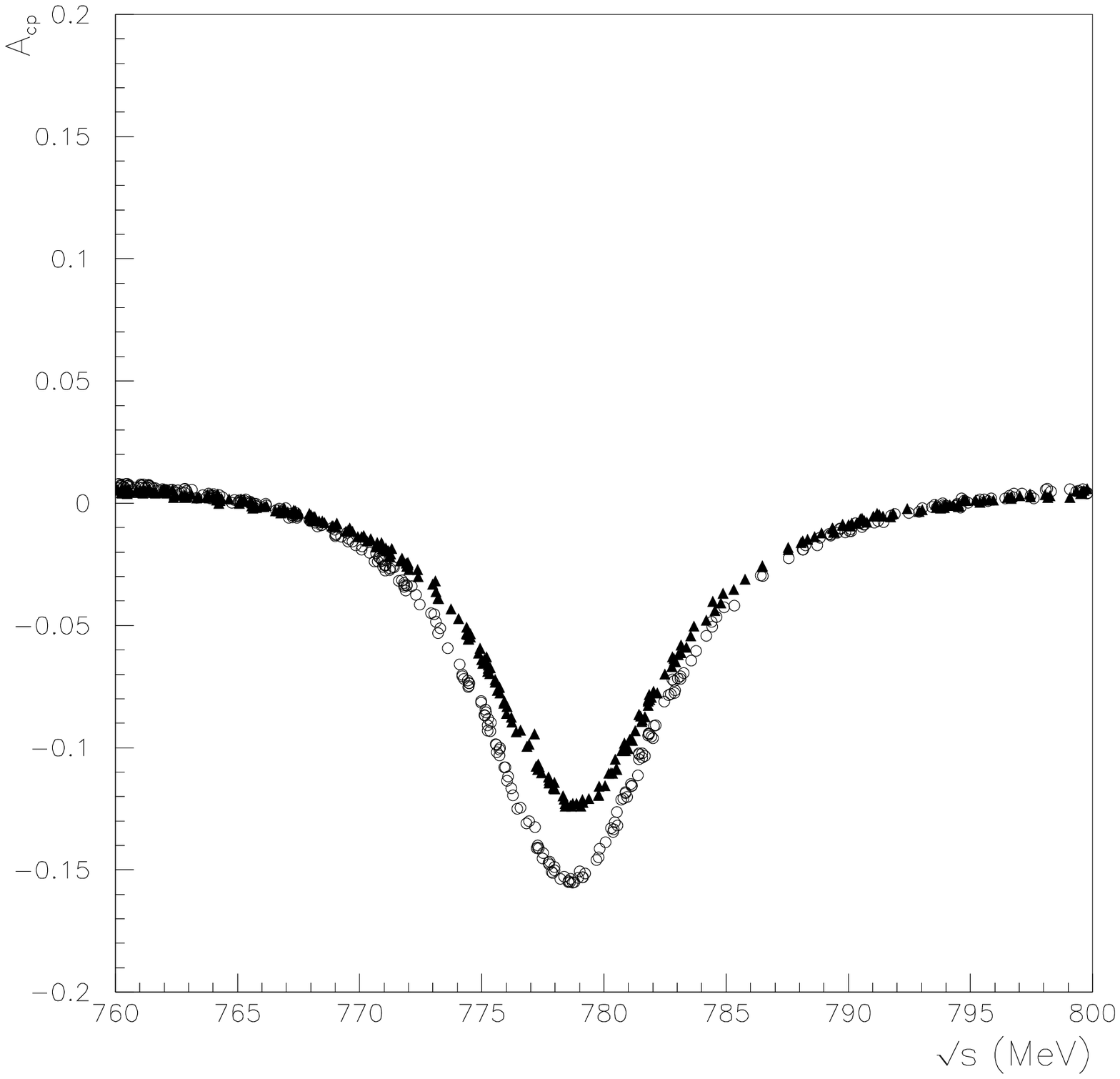}
\caption{ $CP$-violating asymmetry parameter $a_{CP}(m)$, as a function of the
  ${\pi}^+ {\pi}^-$
invariant mass in the vicinity of the $\omega$ mass region for the channel $B^0
\rightarrow \rho^{0}(\omega)
  K^{* 0}$. Parameters  are: ${q^2}/{{m^2}_b} =  0.3$, $N_c^{eff} = 2.84$, $\rho= 0.229,
  \eta=0.325$. Solid triangles up and circles correspond to the BSW and GH form
  factor models respectively.}
\label{fig4}
\end{figure}
\begin{figure}
\centering\includegraphics[height=9.0cm, width = 8.0cm,clip=true]{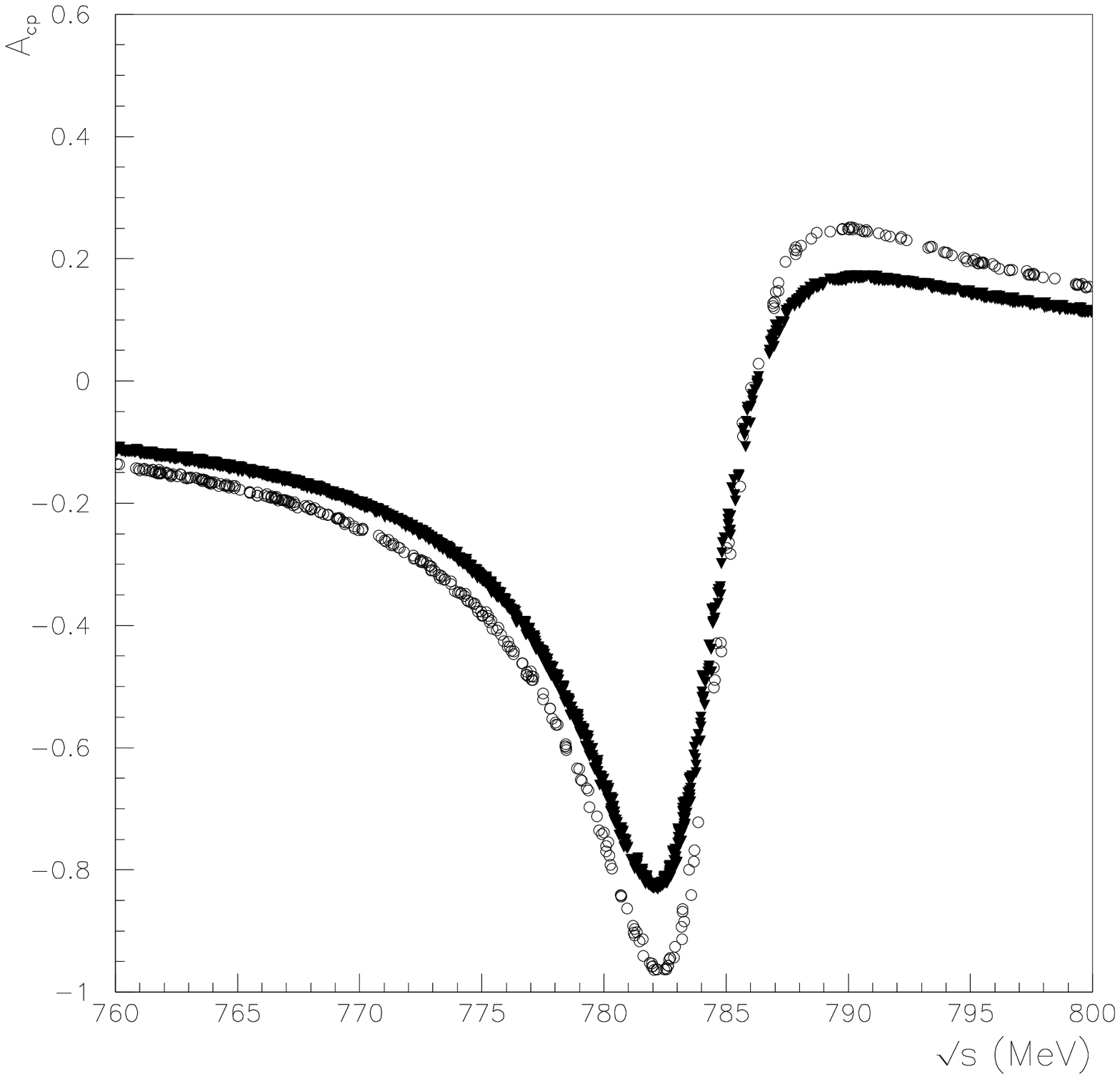}
\caption{ $CP$-violating asymmetry parameter $a_{CP}(m)$, as a function of the
 ${\pi}^+ {\pi}^-$
invariant mass in the vicinity of the $\omega$ mass region for the channel $B^+ \rightarrow \rho^{0}(\omega)
  \rho^+$. Parameters  are: ${q^2}/{{m^2}_b} =  0.3$, $N_c^{eff} = 2.01$,
  $\rho= 0.229,  \eta=0.325$. Solid triangles down and circles 
  correspond to the BSW and GH form
  factor models respectively. }
\label{fig5}
\end{figure}

%

\begin{thebibliography}{99}
\bibitem{Dunietzetal}
I. Dunietz {\it et al}, Phys. Rev. {\bf D43}  (1991) 2193.
\bibitem{Rose}
M.E. Rose, {\it "Elementary theory of angular momentum"}, Dover.
\bibitem{Quinn}
H. Quinn, {\it "Hadronic effects in two-body B decays"}. Lectures at SLAC Summer Institute (1999).
\bibitem{Burasetal} A.J. Buras, Lect. Notes Phys. {\bf 558} (2000) 65, also in `Recent Developments in Quantum Field 
Theory',
Springer Verlag, edited by P. Breitenlohner, D. Maison and J. Wess (Springer-Verleg, Berlin, in press),
 hep-ph/9901409;
 R. Fleischer, Int. J. Mod. Phys. {\bf A12} (1997) 2459, Z. Phys. {\bf C62} (1994) 81, 
Z. Phys. {\bf C58} (1993) 483.
\bibitem{BSW} 
M. Bauer, B. Stech and M. Wirbel, Z. Phys. {\bf C34} (1987) 103; 
M. Wirbel, B. Stech and M. Bauer, Z. Phys.  {\bf C29} (1985) 637.
\bibitem{Ziadetal}
Z.J. Ajaltouni {\it et al}, Eur.Phys.J. {\bf C 29}, 215-233 (2003).
\bibitem{Langacker}
P. Langacker, Phys.Rev. {\bf D20}, 2983 (1979).
\bibitem{Babaretal}
B. Aubert {\it et al} (BaBar collaboration), "Rates, Polarizations and asymmetries in Charmless Vector-Vector B Meson
Decays", hep-ex/0307026;
J. Zhang {\it et al} (Belle collaboration), "Observation of $B^+ \to {\rho}^+ {\rho}^0$ " , hep-ex/0306007.
\bibitem{Thomas}
O. Leitner, X.-H. Guo and A.W. Thomas, Phys. Rev. {\bf D66} (2002) 096008, Phys. Rev. {\bf D63} (2001) 056012.
\bibitem{Rimbault}
C. Rimbault, PhD thesis, report in progress, ''{\it Etude de la violation directe de $CP$ dans la 
d\'esintegration du m\'eson $B$ en deux m\'esons vecteurs non charm\'es. Analyse
du canal $K^{*0} \rho^{0}(\omega)$ dans le cadre de l'experience LHCb.}'';
O. Leitner, PhD Thesis, ''{\it Direct $CP$ violation in $B$ decays including $\rho^{0}-\omega$ mixing and covariant 
light-front dynamics}''.





\end{thebibliography}
\end{document}